\documentstyle[preprint,aps,epsf]{revtex}
\begin{document} 
\title {Exact results for resonating valence bonds states on 2D (narrow) systems}
\author{Moshe Havilio\cite{email}}
\address{Department of Physics, Technion -- Israel Institute of Technology, 
Haifa 32000, Israel}
\date{\today}
\maketitle
\begin{abstract}
It is shown that the problem of calculating spin-spin correlation functions, in the dimers RVB
states, on a possibly diluted 2D square lattice,  can be formulated in terms of a transfer
matrix. The transfer matrix is used for exact numerical calculations of 
spin-spin correlation functions on ladders up to four units wide.
\end{abstract}


In the last few years, the class of Resonating Valence Bonds (RVB) states 
have drawn much attention in connection to 
the long lasting pursuit of the anti-ferromagnetic 2D Heisenberg model ground state 
\cite{ande,assabook}. 
Especially important is the case of spin half, which is considered to be
relevant to high Tc superconductors. 
In this context, simulations had been performed to estimate the 
spin-spin correlation functions in these states\cite{lda};
and particularly the expectation values of the spin half Heisenberg  
Hamiltonian were calculated for variational
considerations. In the  two legged ladder lattice 
the short range RVB states, which are 
in the focus of this article, were found to have 
low energies\cite{nowak}. Despite this interest these simulations had not been
backed by exact results.  This article provides a way to extract
such results, in the case of the 2D dimers RVB states on a (possibly) diluted lattice,
by means of reducing the problem of calculating the spin-spin correlation functions,
to the considerably simpler problem of treating a transfer matrix defined on a 1D lattice.  

 On a general lattice the class of RVB states is defined by \cite{lda}
 \begin{eqnarray}
\mid \! \Psi \rangle=
\sum_{\{\Pi\}}\prod_{\left(\mbox{\boldmath $i$},\mbox{\boldmath $j$}\right) \in \Pi}
{U\left(\mbox{\boldmath $i$},\mbox{\boldmath $j$}\right)}
\left(\uparrow_{\mbox{\boldmath $i$}}\downarrow_{\mbox{\boldmath $j$}} -
\uparrow_{\mbox{\boldmath $j$}}\downarrow_{\mbox{\boldmath $i$}}\right)
\label{rvb states}
\end{eqnarray}
where $\left\{ \Pi \right\}$ are all the possible divisions of  the lattice into pairs of
sites - $ \left({\mbox{\boldmath $i$}}, {\mbox{\boldmath $j$}}\right)$, in which
any site is a member of one and only one pair,  and
 $U\left(\mbox{\boldmath $i$},\mbox{\boldmath $j$}\right)$ is nonnegative. 
For the short-range RVB states $U\left(\mbox{\boldmath $i$},\mbox{\boldmath $j$}\right)$
has a cutoff. We will restrict $\mid\!\Psi\rangle$  to be a 
dimers RVB state on a square lattice, hence
the function $U\left(\mbox{\boldmath $i$},\mbox{\boldmath $j$}\right)$ is given by
\begin{equation}
U\left(\mbox{\boldmath $i$},\mbox{\boldmath $j$}\right)
=\left\{ \begin{array}{lll}
x   & \mbox{if $ \mbox{\boldmath $i$}-\mbox{\boldmath $j$} $} = 
\hat{\mbox{\boldmath $x$}} \\
y   & \mbox{if $\mbox{\boldmath $i$}-\mbox{\boldmath $j$} $} = 
\hat{\mbox{\boldmath $y$}} \\
0   & \mbox{otherwise} 
\end{array} \right. 
\label{di}
\end{equation}
where $\hat{\mbox{\boldmath $x$}}$ and 
$\hat{\mbox{\boldmath $y$}}$ are unit vectors. In this case we can replace $\{\Pi\}$ by 
the dimers coverings of the lattice, denoted by $\left\{ \Delta \right\}$.

The spin-spin correlation function in the  dimers RVB states is defined by
the expectation value\cite{lda,shafir}
\begin{equation}
{\cal S}_{\mbox{\boldmath $i j$}}=\frac{
\langle\Psi\!\mid\mbox{\boldmath $S_{\mbox{\boldmath $i$}}\cdot 
S_{\mbox{\boldmath $j$}}$}\mid \! \Psi\rangle}{
\langle\Psi\!\mid\! \Psi\rangle}
\label{cf}
\end{equation}
where  $ {\mbox{\boldmath $i$}}$ and $ {\mbox{\boldmath $j$}}$ are two sites on the lattice.
Two quantities are evaluated for the calculation of this expectation value.
The norm : 
\begin{eqnarray}
{\cal D} =\langle\Psi\!\mid\! \Psi\rangle &=&
\sum_{\left\{ {\Delta}_{L}, {\Delta}_{R} \right\}}
2^{{\cal N}_{\lambda}}  y^{n_{y}} x^{n_{x}} \nonumber \\ &=&
y^{MK}\sum_{\left\{ {\Delta}_{L}, {\Delta}_{R}\right\}}
2^{{\cal N}_{\lambda}} \eta^{n_{x}} 
\label{dnorm}
\end{eqnarray}
where $\Delta_{L}$ and $\Delta_{R}$ are any 
two dimers coverings of the lattice 
which are placed on each other to create 
$\left\{ {\Delta}_{L}, {\Delta}_{R} \right\} $ which is the ensemble
of all the loops configurations given by overlaps of dimers coverings, this 
ensemble on a - $2 \times 2$ lattice
is depicted in Fig. \ref{fourloops}; $n_{x}$ and $n_{y}$
are the numbers of horizontally and vertically placed dimers respectively
(with $n_{x} + n_{y} = N$, where $N$ is the number of sites in the lattice);
${\cal N}_{\lambda}$ is the number of loops in this 
overlap; and $\eta = \frac{x}{y}$.  And 
\begin{eqnarray}
{\cal C}_{\mbox{\boldmath $i j$}} &=& \pm \frac{4}{3}
\langle\Psi\!\mid\mbox{\boldmath $S_{\mbox{\boldmath $i$}} \cdot 
S_{\mbox{\boldmath $j$}}$}
\mid \! \Psi\rangle \nonumber \\ &=& 
y^{MK}\sum_{\left\{{\Delta}_{L}, {\Delta}_{R}\right\}_{\mbox{\boldmath $i j$}}}
2^{{\cal N}_{\lambda}} \eta^{n_{x}} 
\label{cij}
\end{eqnarray}
where the sign is '$+$' when the two sites are on the same sub-lattice and '$-$' 
otherwise, and
$\left\{{\Delta_{L},\Delta_{R}}\right\}_{\mbox{\boldmath $i j$}}$ is the ensemble
of all the loops configurations in which the   
sites $ {\mbox{\boldmath $i$}}$ and $ {\mbox{\boldmath $j$}}$ are on the same loop -
$\lambda_{\mbox{\boldmath $i j$}}$ 
(this loop is particular to each of the loops configurations). See  Fig. \ref{fourloops} 
for further explanation.

As an introduction,  let us recall the construction of the transfer matrix, 
formulated to solve the dimers problem \cite{lieb}, that is, to calculate the partition function
\begin{equation}
Z = \sum_{\left\{ \Delta \right\}}   y^{n_{y}} x^{n_{x}}. 
\label{zz}
\end{equation} 
In an $M$ rows by $K$ columns square lattice 
the transfer matrix - $V$ is an 
operator defined in a $2^K$ dimensional Hilbert space of the Ising states of $K$
spins half placed in the sites of a $K$ sites row, each in a site.
In a certain dimers configuration, each row in the lattice is represented by an Ising state.
This Ising state denotes the  {\em vertical} dimers which are 
placed on the bonds between this row and the row above it. An up spin in a site 
signifies the presence of a vertical dimer on the bond between this site and the site above it 
(an up-dimer), while a down spin signifies the  absence of an up-dimer on
this bond. We will use the  Fock representation where the spins  are represented by Bosons 
(or Fermions) of spin half and
\begin{equation}
\mid s_1, \: . \:.\: . \:, s_K \rangle \equiv \prod_i^K \left(
\mbox{\boldmath $ c^{\dagger}_{i, s_i} $}\mid \! 0\rangle_i \right)
\label{xxx}
\end{equation}
where $s_i = \uparrow \mbox{or} \downarrow$.

Given an up-dimers configuration 
of a certain row, starting with the first row, $V$ will generate all the permitted dimers 
configurations of the next row, each marked by its up-dimers.  
$V$ is copmposed of two operators :
\begin{eqnarray*}
V = V_2 V_1.
\end{eqnarray*}
The first is
\begin{equation}
V_1=\prod_{i=1}^{K}\sigma^x_i
\label{v1}
\end{equation}
where 
\begin{eqnarray*}
\sigma^{x}_i =\sigma^{+}_i +  \sigma^{-}_i , \; 
\sigma^{+}_i \equiv 
\mbox{\boldmath $ c^{\dagger}_{i \uparrow} c_{i \downarrow}$},\;
\sigma^{-}_i  \equiv \mbox{\boldmath $ c^{\dagger}_{i \downarrow} c_{i \uparrow}$}.
\end{eqnarray*}
This operator reverses all the spins for the next row.
After this operation a vertical or horizontal dimer  can be placed on any up-spin site.
A vertical  dimer is placed by leaving the spin up in the site, and a horizontal dimer, 
between any two adjacent up-spin sites $i$ and $(i+1)$, is placed 
by the operator $\sigma^{-}_i \sigma^{-}_{i+1}$. The operator which places $m$ horizontal
dimers  in arbitrary locations along the row is 
$\frac{1}{m!}\left(\sum_{i=1}^{K}\sigma^{-}_i \sigma^{-}_{i+1}\right)^m$.
An arbitrary number of horizontal
dimers, each accompanied by the weight $\eta$,  are placed by the operator
\begin{equation}
V_2=\exp{\left(\eta\sum_{i=1}^{K}\sigma^{-}_i \sigma^{-}_{i+1}\right)}
\label{v2}
\end{equation}
where  in the case of periodic boundary conditions (which are henceforth assumed) 
$\sigma^{-}_{K+1} \equiv \sigma^{-}_{1}$.
Hence we get (note that in Eq. (\ref{zz})  $n_{x} + n_{y} = MK/2$)  
\begin{equation}
Z = y^{\frac{1}{2}MK} Tr  V^M . 
\label{vv}
\end{equation}

Let us now consider the square of Eq. (\ref{zz}) 
\begin{equation}
Z^2 = \sum_{\left\{ {\Delta}_{L}, {\Delta}_{R} \right\}}
y^{n_{y}} x^{n_{x}} .
\label{dpr}
\end{equation}
This is just summing over loops configurations in whice each site is connected by
dimers of two flavours -  
$L$ and $R$, which are independent of each other. To formulate a transfer 
matrix, we denote  each of the flavoured dimers in any site, 
by a flavoured spin. These spins are propagating  along the columns by the operator
\begin{eqnarray*}
\bar{V} = V_{L}V_{R}
\end{eqnarray*}
where $V_{L}$ and $V_{R}$ are defined in a Hilbert space which is the tensor
product of a $L$ and $R$ 
Ising spaces, and $V_{L}$ ($V_{R}$) effects only the $L$ ($R$) spins in each of the
product states.

After the prepatatory examples above we are ready to tackle the transfer matrix 
represantation of Eq. (\ref{dnorm}), with the additional term of $2^{{\cal N}_{\lambda}}$
for each loops configuration.
Imagine that there are two colours of dimers -  red ($r$) and green ($g$).  
A dimer from each of the colours is distincted by 
an additional flavour index - $\alpha \in \left\{L,R\right\}$. 
We will define the ensemble  $\left\{ \Theta\right\}$ to include all the possible 
coloured-dimers configurations in which each site
in the lattice is connected with two dimers, of identical colours, 
marked by distinct flavour index.  
In $\left\{ \Theta\right\}$ there are exactly $2^{{\cal N}_{\lambda}}$ 
different configurations for 
any loops configuration  in Eq. (\ref{dpr}),
since each of the loops in this overlap can appear in two 
forms - all red, or all green; and we can write
\begin{equation}
{\cal D} = y^{MK}\sum_{\left\{ \Theta \right\}} \eta^{n_{x}}.
\label{tet}
\end{equation}

We denote  the  coloured
dimers in each site by coloured spins. Accordingly we expand further our Hilbert
space to be the $8^K$ dimensional
space of a row of $K$ sites; on each site - $i$ two spin half Bosons, of the flavours 
$L$ and $R$, both  labeled by the {\em same}
colour index - $c_i \in \{r, g\}$. Each of the states in this space is specified by
\begin{eqnarray}
\mid s_{\left(1,L\right)}, s_{\left(1,R \right)},c_1;\: . \:.\: . \: ; 
s_ {\left(K,L\right)}, s_{\left(K,R\right)},c_K\rangle
 \equiv \nonumber \\ \prod_i^K \left(
\mbox{\boldmath $ c^{\dagger}_{i, c_i, L, s_{\left(i,L\right)}} $}
\mbox{\boldmath $ c^{\dagger}_{i, c_i, R, s_{\left(i,R\right)}} $} 
\mid \! 0\rangle_i \right)
\label{yyyz}
\end{eqnarray}

Flipping spins must now include the two colour options so
\begin{equation}
\tilde{V}_1=\prod_{i=1}^{K}\left(\sigma^{x}_{i, r, L} \sigma^{x}_{i, r, R} +
\sigma^{x}_{i, g, L} \sigma^{x}_{i, g, R}\right)
\label{tv1}
\end{equation}
where for example
\begin{eqnarray*}
\sigma^{x}_{i,r,R} \equiv \sigma^{+}_{i, r, R} + \sigma^{-}_{i, r, R} \equiv
\mbox{\boldmath $ c^{\dagger}_{i, r, R, \uparrow} c_{i,r,R, \downarrow}$} +
\mbox{\boldmath $ c^{\dagger}_{i, r, R, \downarrow} c_{i,r,R, \uparrow}$}
\end{eqnarray*} 
Horizontal dimers are placed by lowering spins of the same flavour and colour by four 
operators of the form 
\begin{equation}
\tilde{V}^{c, \alpha}_2=\exp{\left(\eta\sum_{i=1}^{K}\sigma^{-}_{i, c, \alpha}
\sigma^{-}_{i+1, c, \alpha}\right)} 
\label{trv2}
\end{equation}
where $\alpha = L, R$ and $c = r, g$, so now
\begin{equation}	
\tilde{V}_2=\prod_{\begin{array}{cc}
\alpha=L,R\\
c=r,g\\
\end{array}}
\tilde{V}^{c, \alpha}_2.
\label{tv2}
\end{equation}

But this is not all. The operator defined in Eq.(\ref{tv1}) preserves the colour of a column, 
hence imposing the mono-colour 
propagation of a loop with a vertical dimer. 
But if a site is not connected by a vertical dimer from
'below', its pair of dimers should be of either colours, as their are
two colour options for the loop  passing through this site.
Since the absence of vertical dimers is denoted by two down spins,
the additional colour flipping options  are provided by the operator
\begin{eqnarray}
\tilde{V}_0 & = & 
\prod_{i=1}^{K}\left( 1 +  \mbox{\boldmath $ c^{\dagger}_{i,r,L,\downarrow}
 c^{\dagger}_{i,r,R,\downarrow}c_{i,g,L,\downarrow}
c_{i,g,R,\downarrow}$} \right.  \nonumber  \\ & & + \left.   
 \mbox{\boldmath $ c^{\dagger}_{i,g,L,\downarrow}
 c^{\dagger}_{i,g,R,\downarrow}c_{i,r,L,\downarrow}
c_{i,r,R,\downarrow}$}\right) .
\label{v3}
\end{eqnarray}
Combining equations (\ref{tv1}),(\ref{tv2}), and (\ref{v3}), the transfer operator is
\begin{equation}
\tilde{V}=\tilde{V}_2 \tilde{V}_1 \tilde{V}_0
\label{tv}
\end{equation}
and 
\begin{equation}
{\cal D} = y^{MK} Tr {\tilde{V}}^M.
\label{dtrv}
\end{equation}
On narrow lattices, such as the two legged ladder, 
it is possible to extend this treatment to longer range, like
the first and second, or the first and third nearest neighbours, RVB states \cite{hhh}. 
This extension, although straightforward, requires a more complicated transfer matrix, 
which is usualy larger in its dimesion (for a given lattice width). 

The transfer matrix formulation will be concluded in adding, that a disconnected site   
(a static hole) $ {\mbox{\boldmath $i$}}$, can be placed by positioning the operator
\begin{equation}
\tilde{V}^{d}_{\mbox{\boldmath $i$}}= \frac{1}{y} \;
\sigma^{-}_{{\mbox{\boldmath $i$}},r, L} 
\sigma^{-}_{{\mbox{\boldmath $i$}},r, R}
\label{mon}
\end{equation}
to the left of $\tilde{V}_1$, when propagating in to $ {\mbox{\boldmath $i$}}$'s row. This
operator colours the disconnected site in red.  
A class of variational states, that include annealed holes, which obey Fermi statistics, 
will be treated elsewhere, using a similar technique.
 
In order to calculate ${\cal C}_{\mbox{\boldmath $i j$}}$ defined in Eq. (\ref{cij}),
 we will calculate 
the quantity  ${\cal Y}_{\mbox{\boldmath $i j$}}$, defined by
\begin{equation}
{\cal Y}_{\mbox{\boldmath $i j$}}=
y^{MK}\sum_{\left\{\Upsilon\right\}_{\mbox{\boldmath $i j$}}}
2^{{\cal N}_{\lambda}} \eta^{n_{x}} 
\label{upij}
\end{equation}
where $\left\{\Upsilon\right\}_{\mbox{\boldmath $i j$}}$ is the  ensemble of all the loops
configurations in which the   
sites $ {\mbox{\boldmath $i$}}$ and $ {\mbox{\boldmath $j$}}$ are {\em not} on the same
loop. Some of these configurations are
easily counted, by fixing the two pairs of dimers, placed on the two sites,  
to be of distinct colours, or by  calculating the colour-colour correlation function
\begin{equation}
{\cal Y}^{r,g}_{\mbox{\boldmath $i j$}}=
y^{MK} Tr \left(\tilde{V}^{\left(M - p\right)}\mbox{\boldmath $n$}^{r}_
{\mbox{\boldmath $j$},\alpha} 
\tilde{V}^{p}{\mbox{\boldmath $n$}}^{g}_
{\mbox{\boldmath $i$},\alpha}\right)
\label{yyy}
\end{equation}
where the two sites are $p$ rows apart, and
\begin{eqnarray*} 
 \mbox{\boldmath $n$}^{c}_{{\mbox{\boldmath $i$}},\alpha}  =
\mbox{\boldmath $n$}^{c}_{{\mbox{\boldmath $i$}}, \alpha, \uparrow} + \;
\mbox{\boldmath $n$}^{c}_{{\mbox{\boldmath $i$}}, \alpha, \downarrow}
\end{eqnarray*}  
(the flavour index - $\alpha$ is arbitrary). 
In Eq. (\ref{yyy}) we calculate the 
contributions to ${\cal Y}_{\mbox{\boldmath $i j$}}$ from configurations in which  
$\lambda_{\mbox{\boldmath $i$}}$ is a green loop and $\lambda_{\mbox{\boldmath $j$}}$
is a red loop. Each of the other configurations in which $\lambda_{\mbox{\boldmath $i$}}
\neq \lambda_{\mbox{\boldmath $j$}}$ (that is, loops configurations which {\em do 
not} contribute to  
${\cal Y}^{r,g}_{\mbox{\boldmath $i j$}}$, but must be counted in 
${\cal Y}_{\mbox{\boldmath $i j$}}$) has its duplicate in one and only one
of the configurations which do contribute to 
${\cal Y}^{r,g}_{\mbox{\boldmath $i j$}}$, since each of the two loops
may apear in either of the two colours. 
Because there are four colours possibilities for the two loops,
we conclude that
\begin{equation}
{\cal Y}_{\mbox{\boldmath $i j$}} = 4\cdot{\cal Y}^{r,g}_{\mbox{\boldmath $i j$}}.
\label{rgyy}
\end{equation}

Finally, since the sum for the norm in Eq. (\ref{dnorm}) contains all the loops configurations, 
those in which the two sites are on the same loop, and those in which they are not, we get
\begin{equation}
{\cal C}_{\mbox{\boldmath $i j$}} = {\cal D} - {\cal Y}_{\mbox{\boldmath $i j$}}.
\label{ ccc}
\end{equation}

The spin-spin correlation functions
were numericaly calculated, in the isotropic dimers RVB state ($y = \eta = 1$),
on non-diluted ladders of the sizes $2, 3, 4 \times 40$, 
with vertical (along the ladders legs) periodic boundary conditions, and horizontal 
periodic boundary conditions in the case of $4 \times 40$ lattice. 
The results are summerized in table \ref{results}.

In conclusion, using the transfer matrix method it was shown, that it is possible
to reduce the complexity of calculating spin-spin correlation functions 
in dimers RVB states on possibly diluted 2D lattices, 
to the complexity of a 1D quantum many-body problem. This simplification
permited exact calculation of correlation functions on narrow lattices.
We suggest that these kind of calculations are possible when longer range
RVB states are considered.  

The author thanks Assa Auerbach for essential guidance. This work was 
supported by a grant from the Israely Academy of Science, and the Foundation for
Promotion of Research at Technion.

\begin{table}
\begin{tabular}{|c|c|c|c|} \hline
width           &2              & 3                       & 4 \\ \hline
(a)             &$0.286139$     &$0.328752$       &$0.246393$ \\
(b)             &               &$0.277263$       &           \\ 
(c)             &$0.539779$     &$0.344843$       &$0.395096$ \\ \hline
$\mid\!E\!\mid$ &$0.556029$     &$0.656432$       &$0.641489$ \\ \hline
(a)             &$0.696652$ \cite{good}
                                &$0.735272$ \cite{bad}
                                                  &$0.723966$  \\
(b)             &               &$0.780763$       &            \\ \hline
\end{tabular}
\vspace{.5in}
\caption{Results of the one lattice unit spin-spin correlation functions, 
energies, and correlation lengthes, for the isotropic dimers RVB state on
non-diluted ladders of the width 2, 3, and 4 lattice units. 
The results in the second row are  for the absolute values of the  one lattice
unit correlations:  (a) in the vertical case along one of the side legs of the ladders; 
(b)  in the vertical case along the middle  leg of  a $3 \times 40 $ ladder;
and (c) in the horizontal case.
The results in the third row are  for the expactation values of the isotropic Heisenberg  
Hamiltonian for each of the lattices. The  expactation values are in units of $J/site$.
The results in the fourth row are for the correlation lengthes along the (a) side legs 
of the ladders, 
and (b) middle leg of a $3 \times 40 $ ladder.
The  results for the correlation lengthes are according to 
the values of the correlation function in  distances of five and six
lattice units along the lattice. }
\label{results}
\end{table}
\newpage
\newpage

\begin{figure}
\setlength{\unitlength}{0.00500000in}%
\begingroup\makeatletter\ifx\SetFigFont\undefined
\def\x#1#2#3#4#5#6#7\relax{\def\x{#1#2#3#4#5#6}}%
\expandafter\x\fmtname xxxxxx\relax \def\y{splain}%
\ifx\x\y   
\gdef\SetFigFont#1#2#3{%
  \ifnum #1<17\tiny\else \ifnum #1<20\small\else
  \ifnum #1<24\normalsize\else \ifnum #1<29\large\else
  \ifnum #1<34\Large\else \ifnum #1<41\LARGE\else
     \huge\fi\fi\fi\fi\fi\fi
  \csname #3\endcsname}%
\else
\gdef\SetFigFont#1#2#3{\begingroup
  \count@#1\relax \ifnum 25<\count@\count@25\fi
  \def\x{\endgroup\@setsize\SetFigFont{#2pt}}%
  \expandafter\x
    \csname \romannumeral\the\count@ pt\expandafter\endcsname
    \csname @\romannumeral\the\count@ pt\endcsname
  \csname #3\endcsname}%
\fi
\fi\endgroup
\begin{picture}(965,355)(38,86)
\thicklines
\put(560,360){\oval( 80,160)[tl]}
\put(560,360){\oval( 80,160)[bl]}
\put(880,360){\oval( 80,160)[br]}
\put(880,360){\oval( 80,160)[tr]}
\put(560,160){\oval( 20, 80)[tl]}
\put(560,160){\oval( 20, 80)[bl]}
\put(560,160){\oval( 20, 80)[br]}
\put(560,160){\oval( 20, 80)[tr]}
\put(640,160){\oval( 20, 80)[tl]}
\put(640,160){\oval( 20, 80)[bl]}
\put(640,160){\oval( 20, 80)[br]}
\put(640,160){\oval( 20, 80)[tr]}
\put(840,200){\oval( 80, 20)[tr]}
\put(840,200){\oval( 80, 20)[tl]}
\put(840,200){\oval( 80, 20)[bl]}
\put(840,200){\oval( 80, 20)[br]}
\put(840,120){\oval( 80, 20)[bl]}
\put(840,120){\oval( 80, 20)[br]}
\put(840,125){\oval( 80, 20)[tr]}
\put(840,125){\oval( 80, 20)[tl]}
\put( 80,360){\oval( 80,160)[tl]}
\put( 80,360){\oval( 80,160)[bl]}
\put( 80,320){\circle*{10}}
\put(160,400){\circle*{10}}
\put(160,320){\circle*{10}}
\put(320,400){\circle*{10}}
\put(320,320){\circle*{10}}
\put(400,400){\circle*{10}}
\put(400,320){\circle*{10}}
\put(560,400){\circle*{10}}
\put(560,320){\circle*{10}}
\put(640,400){\circle*{10}}
\put(640,320){\circle*{10}}
\put(800,400){\circle*{10}}
\put(800,320){\circle*{10}}
\put(880,400){\circle*{10}}
\put(880,320){\circle*{10}}
\put(480,360){\circle*{10}}
\put( 80,200){\circle*{10}}
\put( 81,121){\circle*{10}}
\put(161,201){\circle*{10}}
\put(160,121){\circle*{10}}
\put(321,201){\circle*{10}}
\put(321,121){\circle*{10}}
\put(401,201){\circle*{10}}
\put(401,121){\circle*{10}}
\put(561,201){\circle*{10}}
\put(561,121){\circle*{10}}
\put(641,201){\circle*{10}}
\put(641,121){\circle*{10}}
\put(801,201){\circle*{10}}
\put(801,121){\circle*{10}}
\put(881,201){\circle*{10}}
\put(881,121){\circle*{10}}
\put( 80,400){\circle*{10}}
\put(240,370){\line( 0,-1){ 20}}
\put(230,360){\line( 1, 0){ 20}}
\put(710,360){\line( 1, 0){ 20}}
\put(720,370){\line( 0,-1){ 20}}
\put( 80,315){\line( 0, 1){ 80}}
\put(400,360){\oval( 80,160)[br]}
\put(400,360){\oval( 80,160)[tr]}
\put(160,320){\line( 0, 1){ 80}}
\put( 80, 95){\makebox(0,0)[lb]{\smash{\SetFigFont{8}{9.6}{bf}j}}}
\put(320,400){\line( 1, 0){ 80}}
\put(320,320){\line( 1, 0){ 80}}
\put(560,315){\line( 0, 1){ 80}}
\put(640,315){\line( 0, 1){ 80}}
\put(800,400){\line( 1, 0){ 80}}
\put(800,320){\line( 1, 0){ 80}}
\put(960,370){\line( 1, 0){ 40}}
\put(960,355){\line( 0, 1){  0}}
\put(960,350){\line( 1, 0){ 40}}
\put( 80,120){\line( 0, 1){ 80}}
\put(160,125){\line( 0, 1){ 80}}
\put(320,115){\line( 0, 1){ 80}}
\put(400,120){\line( 0, 1){ 80}}
\put( 80,200){\line( 1, 0){ 80}}
\put( 80,120){\line( 1, 0){ 80}}
\put(315,200){\line( 1, 0){ 80}}
\put(325,120){\line( 1, 0){ 80}}
\put(230,160){\line( 1, 0){ 20}}
\put(480,170){\line( 0,-1){ 20}}
\put(240,170){\line( 0,-1){ 20}}
\put(470,160){\line( 1, 0){ 20}}
\put(710,160){\line( 1, 0){ 20}}
\put(720,170){\line( 0,-1){ 20}}
\put( 85,360){\makebox(0,0)[lb]{\smash{\SetFigFont{8}{9.6}{bf}L}}}
\put(165,360){\makebox(0,0)[lb]{\smash{\SetFigFont{8}{9.6}{bf}L}}}
\put(355,410){\makebox(0,0)[lb]{\smash{\SetFigFont{8}{9.6}{bf}L}}}
\put(355,330){\makebox(0,0)[lb]{\smash{\SetFigFont{8}{9.6}{bf}L}}}
\put(355,210){\makebox(0,0)[lb]{\smash{\SetFigFont{8}{9.6}{bf}L}}}
\put(360, 95){\makebox(0,0)[lb]{\smash{\SetFigFont{8}{9.6}{bf}L}}}
\put(610,155){\makebox(0,0)[lb]{\smash{\SetFigFont{8}{9.6}{bf}L}}}
\put(835,170){\makebox(0,0)[lb]{\smash{\SetFigFont{8}{9.6}{bf}L}}}
\put(835, 90){\makebox(0,0)[lb]{\smash{\SetFigFont{8}{9.6}{bf}L}}}
\put(565,360){\makebox(0,0)[lb]{\smash{\SetFigFont{8}{9.6}{bf}R}}}
\put(645,360){\makebox(0,0)[lb]{\smash{\SetFigFont{8}{9.6}{bf}R}}}
\put(840,330){\makebox(0,0)[lb]{\smash{\SetFigFont{8}{9.6}{bf}R}}}
\put(835,410){\makebox(0,0)[lb]{\smash{\SetFigFont{8}{9.6}{bf}R}}}
\put(835,215){\makebox(0,0)[lb]{\smash{\SetFigFont{8}{9.6}{bf}R}}}
\put(835,140){\makebox(0,0)[lb]{\smash{\SetFigFont{8}{9.6}{bf}R}}}
\put(575,155){\makebox(0,0)[lb]{\smash{\SetFigFont{8}{9.6}{bf}R}}}
\put(655,155){\makebox(0,0)[lb]{\smash{\SetFigFont{8}{9.6}{bf}R}}}
\put(405,155){\makebox(0,0)[lb]{\smash{\SetFigFont{8}{9.6}{bf}R}}}
\put(115,210){\makebox(0,0)[lb]{\smash{\SetFigFont{8}{9.6}{bf}R}}}
\put(115, 95){\makebox(0,0)[lb]{\smash{\SetFigFont{8}{9.6}{bf}R}}}
\put(530,155){\makebox(0,0)[lb]{\smash{\SetFigFont{8}{9.6}{bf}L}}}
\put(170,155){\makebox(0,0)[lb]{\smash{\SetFigFont{8}{9.6}{bf}L}}}
\put( 60,155){\makebox(0,0)[lb]{\smash{\SetFigFont{8}{9.6}{bf}L}}}
\put(295,155){\makebox(0,0)[lb]{\smash{\SetFigFont{8}{9.6}{bf}R}}}
\put( 80,210){\makebox(0,0)[lb]{\smash{\SetFigFont{8}{9.6}{bf}i}}}
\end{picture}
\vspace{.5in}
\caption{The four possible loops configurations, made by overlaps of dimers coverings,  
on a - $2 \times 2$ lattice. The dimers  
configurations from the {\bf L} side are placed on the dimers configurations from the 
{\bf R} side. In the sum of  Eq. (5), for the two sites marked by  
$\bf i$ and $\bf j$, the first (from left to right) three loops 
configurations have to be included.} 
\label{fourloops}  	 
\end{figure}


\begin{thebibliography}{99}
\bibitem{bad} In the case of 3 units width lattice, the correlation function is slightly
wavy, hence a slightly different result  may apear, 
if the correlation length is to be calculeted
by the values of the correlation function in different distances.
\end{thebibliography}
\end{document}